\documentstyle[prl,aps,twocolumn,epsf]{revtex}

\begin{document}
\draft

\twocolumn[
\hsize\textwidth\columnwidth\hsize\csname @twocolumnfalse\endcsname

\title{
Localization effects in disordered Kondo lattices
}
\author{
E. Miranda$^a$ and V. Dobrosavljevi\'{c}$^b$
} 
\address{$^a$Instituto de F\'{\i}sica,
Universidade Estadual de Campinas, CEP
13083-970, Campinas, SP, Brazil\\
$^b$
Department of Physics and 
National High Magnetic Field Laboratory,\\
Florida State University, 
Tallahassee, Florida 32306.
}

\maketitle

\begin{abstract}

We investigate the role of localization effects in the Kondo disorder
mechanism for non-Fermi liquid behavior in disordered Kondo lattices.
We find that the distribution of Kondo temperatures is strongly
affected by fluctuations of the conduction electron density of states,
a feature neglected in the previous treatment. For moderate disorder,
the self-consistent distribution of Kondo temperatures flows to a
universal log-normal form, {\it irrespective} of the form of the bare
disorder distribution. For sufficient disorder, the system enters a
Griffiths phase with diverging thermodynamic responses.

\end{abstract}

\vspace*{1cm}
]

\begin{flushleft}

$^a$ Corresponding author.
\end{flushleft}


The question of the origin of Non-Fermi liquid (NFL) behavior in
metals remains unsolved. The issue has become particularly intriguing
in the case of f-electron materials, where it has elicited such
diverse explanations as the proximity to a quantum critical point,
exotic impurity models or disorder-driven mechanisms\cite{nflgeneral}.
The present authors have emphasized the possibility of explaining the
singular behavior of these systems by considering that disorder leads
to a wide distribution of Kondo temperatures\cite{TKdist,bernal,us}.
In a series of previous papers\cite{us}, we demonstrated that such
disorder effects can explain not only the thermodynamics, but also the
anomalous transport in these systems.  In this case, very low $T_K$
spins remain unquenched at low temperatures and lead to the anomalous
NFL behavior.  The model has been most thoroughly investigated in the
alloys UCu$_{5-x}$Pd$_x$\cite{nfl_ucupd,bernal,doug}, where its
predictions have been quite successful in explaining the available
data.

We formulated a theory appropriate for concentrated magnetic
impurities, able to describe the coherence effects in the clean
limit. We showed that correlation effects strongly enhance any
extrinsic disorder, generating a broad distribution of Kondo
temperatures. The approach used was based on the dynamical mean field
theory (DMFT) of correlations and disorder\cite{dinf}. However, the
DMFT is unable to accomodate localization effects, in that it treats
conduction electron disorder ``on the average'', at the CPA
level. Therefore, an outstanding issue that needs to be addressed is
the role of fluctuations in the local conduction electron density of
states. Indeed, local Kondo temperatures are given by $T_K = D
e^{-1/\rho J}$ and fluctuations of the conduction electron density of
states $\rho$ should be at least equally important in determining the
distribution of Kondo temperatures. The goal of the present study is
to incorporate such Anderson localization effects into our DMFT.

Our approach has been used before to study the Mott-Anderson
transition in a disordered Hubbard model\cite{mott-and}. In it, the
correlation aspects of the problem are taken into account in a DMFT
fashion, but we also allow for {\em spatial variations} of the DMFT
order parameter in order to accommodate Anderson localization
effects. Such approach has been dubbed ``statistical mean field
theory'' (SMFT)\cite{mott-and}.

We concentrate on the disordered Anderson lattice model given by the
Hamiltonian
\begin{eqnarray}
H&=&\sum_{ij\sigma} ( -t_{ij} + \varepsilon_i \delta_{ij})
c^{\dagger}_{i,\sigma}c_{j,\sigma} + \sum\limits_{j\sigma} E^f_j
f^{\dagger}_{j\sigma} f^{\phantom{{\dagger}}}_{j\sigma} \nonumber \\
&+& \sum\limits_{j\sigma} V_j (c^{\dagger}_{j\sigma} f^{\phantom{{\dagger}}%
}_{j\sigma} + {\rm H. c.} ) + U\sum_{i}f^{\dagger}_{i,\uparrow}f^{%
\phantom{\dagger}}_{i,\uparrow} f^{\dagger}_{i,\downarrow}f^{%
\phantom{\dagger}}_{i,\downarrow},  \label{hammy}
\end{eqnarray}
where, in principle, we allow for random c- and f-site energies
($\epsilon_i$ and $E^f_j$) and hybridization matrix elements $V_j$.

The SMFT is considerably simplified when formulated on a Bethe lattice
of coordination $z$ (cf. Ref.~\cite{abouchacra}). We are then led to
solve a set of stochastic equations by sampling. The equations for the
disordered Anderson lattice read\cite{mott-and}
\begin{mathletters}
\label{sdmft}
\begin{eqnarray}
G_{cj}^{(i)(-1)}(\omega ) &=&\omega -\epsilon
_{j}-\sum_{k=1}^{z-1}t_{jk}^{2}G_{ck}^{(j)}(\omega ) \nonumber \\
&-&\frac{V_{j}^{2}}{\omega -E_{j}^{f}-\Sigma _{fj}(\omega )};  \label{recur}
\\
S_{{\rm eff}}^{(j)} &=&\sum_{\sigma }\int_{o}^{\beta }d\tau \int_{o}^{\beta
}d\tau ^{\prime }f_{j,\sigma }^{\dagger }(\tau )\nonumber \\
&&\left[ \delta (\tau -\tau ^{\prime })\left( \partial _{\tau
}+E_{j}^{f}\right)
+\Delta _{j}(\tau -\tau ^{\prime })\right] f_{j,\sigma
}(\tau ^{\prime })  \nonumber \\
&+&U \sum_{\sigma}\int_{o}^{\beta }d\tau 
f^{\dagger}_{i,\uparrow}(\tau)f^{\phantom{\dagger}}_{i,\uparrow}(\tau)
f^{\dagger}_{i,\downarrow}(\tau)f^{\phantom{\dagger}}_{i,\downarrow}(\tau);
\label{seff} \\
\Delta _{j}(\omega ) &=&\frac{V_{j}^{2}}{\omega -\epsilon
_{j}-\sum_{k=1}^{z-1}t_{jk}^{2}G_{ck}^{(j)}(\omega )}.  \label{hybrid}
\end{eqnarray}
\end{mathletters}
Here, $G_{cj}^{(i)}(\omega )$ is the c-electron Green's function on
site $j$ with the nearest neighbor site $i$ removed and $\Sigma
_{fj}{(\omega )} $ is the f-electron self-energy calculated from the
effective action (\ref {seff})\cite{mott-and}. We calculate the
self-energy with the large-N mean-field theory at $T=0$ and $U\to
\infty $.

\begin{figure}[h]
\epsfxsize=3.1in 
\centerline{\epsfbox{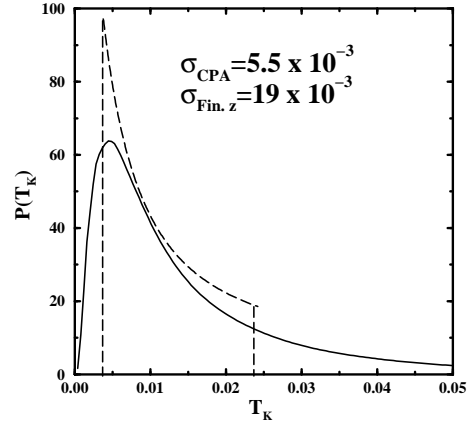}}
\caption{Comparison between $P(T_K)$ with (full line) and without
(dashed line) localization effects. The latter shows an enhanced
standard deviation $\sigma$. This is for a uniform V-distribution with
$<V> = 0.45$ and width $W_V=0.1$, $\mu = -0.1$, $E_f =-1$, in units of
$t$ and $z=3$.
\label{fig1}}
\end{figure}

We briefly summarize our results. We will only discuss the effect of
f-disorder ($E_f$ and $V$). We have found that the inclusion of
localization effects can significantly enhance the width of the
distribution of Kondo temperatures as compared with the CPA results
previously obtained\cite{us}. This effect is depicted in
Fig.~(\ref{fig1}), where $P(T_{K})$ obtained by the present method is
compared with $P(T_{K})$ obtained within the dynamical mean field
theory\cite{us}.

\begin{figure}[h]
\epsfxsize=3.1in 
\centerline{\epsfbox{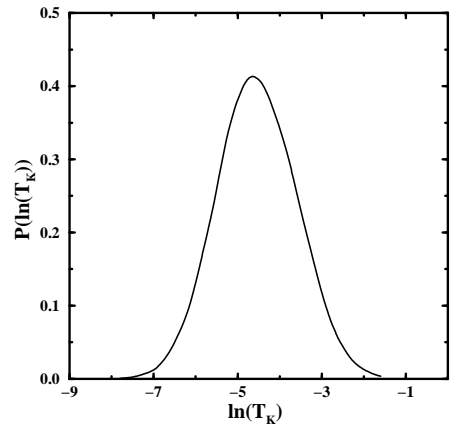}}
\caption{ Universal form of the distribution of $T_K$'s with
localization effects. The bare distribution of $V$'s is constant, but
the self-consistent $P(T_K)$ is log-normal. Same parameters as in
Fig.~(\ref{fig1}).
\label{fig2}}
\end{figure}

Furthermore, irrespective of the bare distribution of physical
parameters, the self-consistent $P(T_{K}) $ {\em flows to a universal
log-normal distribution}, for weak to intermediate disorder. This is
illustrated in Fig.~(\ref{fig2}), where a bare uniform distribution of
$V$'s was used.

\begin{figure}[h]
\epsfxsize=3.1in 
\centerline{\epsfbox{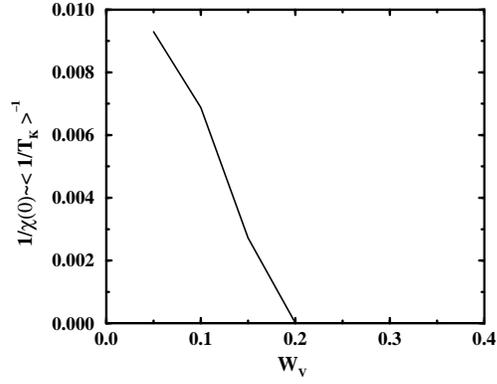}}
\caption{Inverse $T=0$ magnetic susceptibility $\chi(0)$ within the
SMFT for various disorder strengths. Above a threshold disorder
($W_V\sim 0.2$), $\chi(0)\to\infty$. Same parameters as in
Fig.~(\ref{fig1}).
\label{fig3}}
\end{figure}

Finally, for a range of disorder strengths, we have found a Griffiths
phase characterized by non-Fermi liquid behavior with diverging
magnetic susceptibility and linear specific heat coefficient. This is
shown if Fig.~(\ref{fig3}).

In summary, we have explored the effect of fluctuations in the
conduction electron density of states (localization effects) in the
distribution of Kondo temperatures in the Kondo disorder model. These
were shown to play a role that is at least as important as the
fluctuation of Kondo coupling constants considered before.

\begin{references}

\bibitem{nflgeneral} See, e. g., the volume J. Phys.:Cond. Matter {\bf
8} (1996).   

\bibitem{TKdist} V. Dobrosavljevi\'{c}, T. R. Kirkpatrick, and
G. Kotliar, Phys.  Rev. Lett. {\bf 69}, 1113 (1992).

\bibitem{bernal} O. O. Bernal {\it et. al.}, Phys. Rev. Lett. {\bf
75}, 2023, (1995).

\bibitem{us} E. Miranda, V. Dobrosavljevi\'{c}, and G. Kotliar,
J. Phys.: Cond. Matter {\bf 8}, 9871 (1996), Phys. Rev. Lett. {\bf
78}, 290 (1997), Physica B {\bf 230}, 569 (1997).

\bibitem{nfl_ucupd} B. Andraka and G. R. Stewart, Phys. Rev. B {\bf
47}, 3208 (1993); M. C. Aronson {\it et. al.}, Phys. Rev. Lett. {\bf
75}, 725 (1995).

\bibitem{doug} D. E. MacLaughlin {\it et al.}, cond-mat/9804329.

\bibitem{dinf} For a review see, A. Georges {\it et al.},
Rev. Mod. Phys., {\bf 68}, 13 (1996).

\bibitem{mott-and} V. Dobrosavljevi\'{c} and G. Kotliar,
Phys. Rev. Lett.  {\bf 78}, 3943 (1997);
Phil. Trans. R. Soc. Lond. {\bf A356}, 1 (1998).

\bibitem{abouchacra} R. Abou-Chacra, P. W. Anderson, D. J. Thouless,
J.  Phys. C {\bf 6}, 1734 (1973).

\end{references}
\end{document}